\documentclass{appolb}
\usepackage{epsfig}
\usepackage{subfigure}
\begin{document}
\title{The early thermalization and HBT puzzles at RHIC%
\thanks{Lectures presented at the Jubilee 50th Cracow School of Theoretical Physics, Zakopane, 
June 9--19, 2010}%
}
\author{Wojciech Florkowski
\address{Institute of Physics, Jan Kochanowski University, PL-25-406~Kielce, Poland
\\ and H. Niewodnicza\'nski Institute of Nuclear Physics, \\ Polish Academy of Sciences, PL-31-342 Krak\'ow, Poland}
%\and
%the Name(s) of other Author(s)
%\address{and their affiliation}
}
\maketitle
\begin{abstract}
The early thermalization and HBT puzzles in relativistic heavy-ion collisions studied at RHIC are shortly reviewed. The results of recent hydrodynamic calculations that shed light on these two intriguing issues are presented. In particular, the role of the elliptic flow as a signature of early thermalization is critically examined. 
\end{abstract}
  
\section{Introduction}
\label{sect:intro}

The early thermalization and HBT puzzles in relativistic heavy-ion collisions were addresses simultaneously for the first time by Heinz and Kolb in 2002 \cite{Heinz:2002un} (see also \cite{Hirano:2001eu,Heinz:2001xi}). In view of the first successes of relativistic hydrodynamics applied to describe the hadronic spectra at RHIC \cite{Huovinen:2001cy,Teaney:2001av,Kolb:2002ve}, a short thermalization time scale of less than 1 fm ($c=1$) as well as disagreements between the HBT data and the hydrodynamic predictions for correlation functions represented a serious challenge for theory. In the meantime, a substantial progress in our understanding of those two puzzles has been made. At present, the agreement between the advanced hydrodynamic calculations and the HBT data is much better. However, the thermalization time scales assumed in such calculations remain short, well below  1 fm. This leaves still much place for a discussion of the early thermalization puzzle. The latter is most commonly solved by the assumption that matter formed in the relativistic heavy-ion collisions at RHIC is a strongly interacting quark-gluon plasma (sQGP) \cite{Gyulassy:2004zy,Shuryak:2004cy}.  

As the notion of the ``early thermalization puzzle'' reflects simply our prejudice to accept an extremely short thermalization time (not supported by the available microscopic calculations), the  notion of the ``HBT puzzle'' is much more involved. First of all, the name ``HBT analysis'' is not very much informative. It refers mainly to the study of pion correlation functions (the two-particle distributions of identical pions with small relative momenta), for which the names ``pion interferometry'' or ``pion femtoscopy'' seem to be more adequate. The term ``HBT'' emphasizes the analogy to the well known technique of second order  intensity interferometry developed by Hanbury-Brown and Twiss \cite{HanburyBrown:1954wr,HanburyBrown:1956pf}\index{HBT} to measure stellar angular sizes~\footnote{This analogy is very often misunderstood. The particle momentum correlations are confused with the space–time HBT correlations. Although the two types of the correlations have the common quantum statistical origin, the momentum correlations of identical particles yield the space–time picture of the source, whereas the space–time HBT correlations (dependence of the number of coincident two-photon counts on the distance between two detectors) provide the information on the characteristic relative three-momenta of emitted photons, which gives the angular size of a star without the knowledge of its radius and lifetime \cite{Kopylov:1975rp,Baym:1997ce,Lednicky:2003mq}.}. The measurements of the pion correlation functions are quantified by the values of the ``HBT radii'' that define the range of the correlation functions. In this context, one should remember that the HBT radii $R_{\rm side}$, $R_{\rm out}$, and $R_{\rm long}$ give information about the space-time sizes of the regions where pions are correlated (so called homogeneity lengths \cite{Makhlin:1987gm}) rather than about the actual sizes of the whole system. The persistent differences between the measured and modeled HBT radii were termed the ``(RHIC) HBT puzzle''. 

As was stated above, the two puzzles emerge in the context of rich applications of relativistic hydrodynamics to describe heavy-ion collisions at RHIC. Contrary to the calculations based on hydrodynamics or kinetic theory, the one- and two-particle distributions may be consistently described in terms of simple, blast-wave-type models which assume appropriate physical conditions (temperature, transverse flow) on the properly chosen freeze-out hypersurface \cite{Broniowski:2002wp,Retiere:2003kf}.  The problem remains, however, to validate such freeze-out conditions with the help of hydrodynamic or kinetic models which are regarded as more fundamental frameworks.

More information and details about the HBT analysis and the HBT puzzle the reader may find in the recent reviews \cite{Lisa:2005dd,Lisa:2008gf}. In this paper we discuss the HBT puzzle in the context of very recent hydrodynamic calculations. 

\section{Resolving the HBT puzzle}
\label{sect:HBT}

The common present opinion is that several improvements done in the hydrodynamic models may improve their predictions in such a way that the HBT puzzle is practically eliminated \cite{Broniowski:2008vp,Pratt:2008qv} (discrepancy between the data and theory remains at the level of about 10\%). The discussed improvements include:

\begin{itemize}
\item[i)] the use of a realistic QCD equation of state \cite{Chojnacki:2007jc,Huovinen:2009yb} (instead of a simple bag equation of state), 
\item[ii)] early start of hydrodynamics (with the starting times as short as \mbox{$\tau_{\rm eq} \sim $ 0.10--0.25 fm} \cite{Pratt:2008qv,Bozek:2009ty}),  
\item[iii)] inclusion of the pre-equilibrium flow \cite{Chojnacki:2004ec,Sinyukov:2009xe}, 
\item[iv)] inclusion of the shear viscosity \cite{Pratt:2008qv}, 
\item[v)] accounting for the effects of fluctuations of the initial eccentricity \cite{Socolowski:2004hw} (the use of the participant plane rather than the reaction plane in the calculations of the elliptic flow),
\item[vi)] the use of two-particle methods for the calculation of correlation functions \cite{Kisiel:2006is},
\item[vii)] modified initial conditions (for example, the Gaussian profiles of the energy density in the transverse plane \cite{Broniowski:2008vp}),
\item[viii)] fast freeze-out (for example, the use of a single-freeze-out scenario  \cite{Broniowski:2001we} where the collisions between the emitted hadrons are neglected),
\item[ix)] inclusion of a complete set of hadronic resonances \cite{Kisiel:2006is}.
\end{itemize}

In our opinion the consensus has been reached for the points: i), ii), v), and vi). Moreover, the point ii) may be treated, to some extent, as equivalent to iii) and iv). The points vii)--ix) were proposed and emphasized in Ref. \cite{Broniowski:2008vp}. 

Let us now briefly discuss the physical arguments standing behind the points \mbox{i)--ix)}: The use of the realistic QCD equation of state, e.g., see \cite{Chojnacki:2007jc,Huovinen:2009yb}, which is much stiffer than the bag equation of state with large latent heat leads to less extended outward dimensions, which lowers the $R_{\rm out}/R_{\rm side}$ ratio and makes it more compatible with the experimental data indicating $R_{\rm out}/R_{\rm side} \sim 1$. Similar effects are caused by an earlier start of hydrodynamics or by the inclusion of the pre-equilibrium flow \cite{Chojnacki:2004ec,Sinyukov:2009xe} in standard initial conditions. For example, to avoid problems with choosing the appropriate form of the initial transverse flow  at $\tau =$ 1 fm, one may decide to start the hydrodynamic evolution at an earlier time, $\tau < $ 1 fm, and treat the results at \mbox{$\tau = $ 1 fm} as new initial conditions with transverse flow. 

The presence of shear viscosity also increases the explosiveness of the collision \cite{Teaney:2003kp,Romatschke:2007mq,Pratt:2008sz}. This can be seen by considering viscous corrections to the stress-energy tensor. At the beginning of the evolution the velocity gradient is mainly longitudinal, which affects the stress-energy tensor by increasing the transverse pressure and decreasing the longitudinal pressure. 

It should be emphasized that the solution of the HBT puzzle cannot be reduced only to the correct reproduction of the HBT radii in a certain hydrodynamic calculation. In addition, one should describe well also the transverse-momentum spectra and the elliptic flow of hadrons. In other words, the acceptable solution of the HBT puzzle requires that both the one- and two-particle observables are uniformly reproduced~\footnote{The hydrodynamic predictions make sense in the soft hadronic sector where \mbox{$p_\perp < 1-2$ GeV}. Thus, the HBT puzzle discussed here refers to the soft hadronic observables.}. In this context it is important to establish the most reliable method of the calculation of the elliptic flow in the hydrodynamic calculations.

In Ref. \cite{Andrade:2006yh} the important point was made concerning the inclusion of fluctuations of the eccentricity characterizing initial distributions of matter in the transverse plane. Such fluctuations increase the value of the elliptic flow. The inclusion of the fluctuations in the theoretical calculations helps to reconcile the predictions of the models with the data  \cite{Andrade:2006yh}. It was also shown by the same group \cite{Socolowski:2004hw} that the fluctuations improve the agreement between the HBT data and hydrodynamic calculations.

The fluctuations of the eccentricity may be taken into account in a simple way by fixing centrality class and preparing ``typical hydrodynamic initial conditions'' for this class. This can be achieved, for example, by doing Monte-Carlo simulations of the Glauber model with {\tt GLISSANDO} \cite{Broniowski:2007nz}. In this procedure the fluctuations are included in an averaged way. Definitely, a more basic but also a more time-consuming approach requires that the hydrodynamic evolution is applied after each generation of initial conditions by the appropriate model of the early stage (event-by-event hydrodynamics). Such strategy is used now in Refs. \cite{Werner:2010aa,Schenke:2010rr}. 

Early hydrodynamic calculations of the HBT radii used frequently simple formulas relating $R_{\rm side}$, $R_{\rm out}$, and $R_{\rm long}$ to the space-time extension of the produced system. Nowadays, much more sophisticated methods are applied that try to mimic more closely the experimental situation. The two-particle methods are used in the framework of the event generators, e.g., {\tt THERMINATOR} \cite{Kisiel:2005hn}. The correlation functions are constructed with the help of the emission function and the squared wave functions of the pion pairs \cite{Kisiel:2006is}. If the wave functions include the Coulomb interactions, the Bowler-Sinyukov method \cite{Bowler:1991vx,Sinyukov:1998fc} is used to extract the HBT radii. The explicit calculations using simple parameterizations of freeze-out show that the approach using the standard fitting without the Coulomb corrections  agrees very well with the approach that includes the Coulomb interactions and uses the Bowler-Sinyukov \mbox{method \cite{Kisiel:2006is,Maj:2009ue}}. 

In Ref. \cite{Broniowski:2008vp} a very much successful description of the soft hadronic observables studied at RHIC was achieved in the hydrodynamic approach using modified initial conditions for energy density in the transverse plane. Instead of using the Glauber model in the optical limit, the Monte-Carlo Glauber model \cite{Broniowski:2007nz} was used to determine typical initial conditions (distributions) for each centrality class. Then, the Gaussian fits were performed to those distributions, which were subsequently used as the input for the hydrodynamic calculations. It turns out that the initial Gaussian distributions lead to the freeze-out conditions which reproduce very well soft hadronic observables. 

At present no microscopic explanation of the Gaussian initial conditions is known. They do a good job because steeper profiles of the initial energy density in the transverse plane lead to larger pressure gradients at the center of the fireball and speed up development of the transverse flow. In this sense, the point vii) leads to similar effects as the points ii), iii), and iv). In addition, the initial Gaussian profiles lead to the Hubble flow \cite{Broniowski:2008qk} which is a kind of optimal flow for correct reproduction of the HBT radii in the blast-wave models \cite{Chojnacki:2004ec}.  

The calculations \cite{Broniowski:2008vp} used the concept of a single freeze-out \cite{Broniowski:2001we}. In this approach the hadronic rescattering in the final state is neglected. To check if this assumption makes sense, the authors of \cite{Broniowski:2008vp} made estimates of the number of pion and proton collisions after freeze-out. This number was estimated as follows: pion straight-line trajectories after freeze-out were assumed and the number of collisions was counted, i.e., the encounters with other particles closer than the distance corresponding to the cross section. Averaging over all pions yields the number of these trajectory crossings about 1.5–-1.7 per pion. These estimates indicate that the single-freeze-out scenario may be well accepted for pions. For protons, similar estimates are worse indicating non-negligible effects that may change the proton $v_2$, see our discussion below in Sect. \ref{sect:v2puzzle}.

In the calculations of the HBT radii it is important to include the effects of the resonance decays which make the studied systems larger by about \mbox{1 fm} \cite{Kisiel:2006is}. Most of the publicly available codes that simulate resonance decays use the resonance tables from {\tt SHARE} \cite{Torrieri:2004zz}.

\section{Early thermalization problem}
\label{sect:ET}

Within the concept of a strongly coupled quark-gluon plasma short thermalization times appear in the natural way, as the strongly interacting system equilibrates very fast. However, it is still debatable if the plasma in the early stage is indeed strongly coupled. In the last years many different approaches were used to discuss the problem of very early thermalization and very often such approaches considered the plasma as a weakly interacting system:

\begin{itemize}
\item[i)]  The equilibration problem was studied within the parton cascade model. Initially, this model included only binary collisions but further developments took into account the gluon radiation in the initial and final states \cite{Geiger:1994he}. More recently, the advanced numerical codes have been developed, that stress the role of the multi-particle processes. In Ref. \cite{Xu:2007qn} the production and absorption  \mbox{2 $\leftrightarrow$ 3} processes are taken into account, whereas in Ref.  \cite{Xu:2004gw} the three-particle collisions \mbox{3 $\leftrightarrow$ 3} are studied. Within both approaches the equilibration is claimed to be significantly sped up when compared to the equilibration driven by the binary collisions. 
\item[ii)] One usually assumes that the initial partons are produced by hard or semi-hard interactions of partons in the incident nuclei. In the ``bottom-up'' thermalization scenario \cite{Baier:2000sb}, the initial state is described by the QCD saturation mechanism that is incorporated in the framework of the color glass condensate. Thus, the initial state is dominated by the small $x$ gluons of transverse momentum of order $Q_s$ (one expects $Q_s \sim$ 1 GeV for RHIC, and $Q_s \sim$ 2--3 GeV for the LHC \cite{Baier:2000sb}). The calculations performed within the ``bottom-up'' thermalization scenario \cite{Baier:2000sb}, where the binary and \mbox{2 $\leftrightarrow$ 3} processes are taken into account, give an equilibration time of at least 2.6 fm \cite{Baier:2002bt}, see also \cite{El:2007vg}.
\item[iii)] The equilibration of the system is most commonly understood as an effect of parton rescattering. An interesting phenomenon occurs, however, that the equilibration is speeded-up by instabilities generated in an anisotropic quark-gluon plasma\index{quark-gluon plasma (QGP)!anisotropic} \cite{Mrowczynski:1994xv,Arnold:2004ti}. This is so because the growth of the unstable modes is associated with the isotropization of the momentum distributions, which helps to achieve the full equilibration. 
\item[iv)] It is also argued that the very process of particle production leads to the equilibrium state without any secondary interactions. For instance, Refs. \cite{Bialas:1999zg,Florkowski:2003mm} refer to the Schwinger mechanism\index{Schwinger mechanism} of the particle production in the strong chromoelectric field. This approach explains, however, the equilibration of the transverse momentum only. The approach of Refs.~\cite{Kharzeev:2005iz,Castorina:2007eb}, where the longitudinal momentum is also thermal, evokes the Hawking-Unruh effect.
\item[v)] An interesting physical scenario was also formulated, where the thermalization is not an effect of collisions but a consequence of the chaotic dynamics of the non-Abelian classical color fields, coupled or not to the classical colored particles \cite{Biro:1993qc}.
\end{itemize}
One may summarize different approaches to the early-thermalization problem with the conclusion that newly developed simulations based on perturbative QCD as well as new non-perturbative frameworks may explain the thermalization times of about 1 fm. The problem remains to explain much shorter thermalization times used in the hydrodynamic codes that successfully describe the data. For a more detailed discussion of these points see \cite{Mrowczynski:2005ki}.

\section{Hydrodynamic models with delayed thermalization}

The support for the idea of a very short thermalization/equilibration time at RHIC comes mainly from the observation of the large elliptic flow. This effect has been explained by the hydrodynamic (perfect-fluid) expansion with an early starting time. Recent hydrodynamic calculations challenge this point of view. It turns out that it is possible to connect the hydrodynamic perfect-fluid hydrodynamic evolution with the pre-equilibrium stage that is not completely thermalized. The results of such calculations are compatible with the data as we discuss below. 

\subsection{Early free-streaming stage}
\label{sect:model_1}

An approximation to the early-stage dynamics in relativistic heavy-ion collisions consisting of the free streaming (FS) of partons followed by a sudden equilibration (SE) to a thermalized  phase which subsequently undergoes a hydrodynamic evolution was proposed by Kolb, Sollfrank, and Heinz in \cite{Kolb:2000sd}. In the meantime, this approximation has been frequently considered in the modeling of the early stages of the nuclear high-energy collisions in the context of equilibration. In particular, it has been considered in an investigation of the isotropization problem by Jas and Mrowczynski \cite{Jas:2007rw}, as well as elaborated in the context of the early development of flow by Sinyukov, Gyulassy, Karpenko, and Nazarenko \cite{Sinyukov:2006dw,Gyulassy:2007zz}.

\begin{figure}[tb]
\begin{center}
\includegraphics[angle=0,width=0.65 \textwidth]{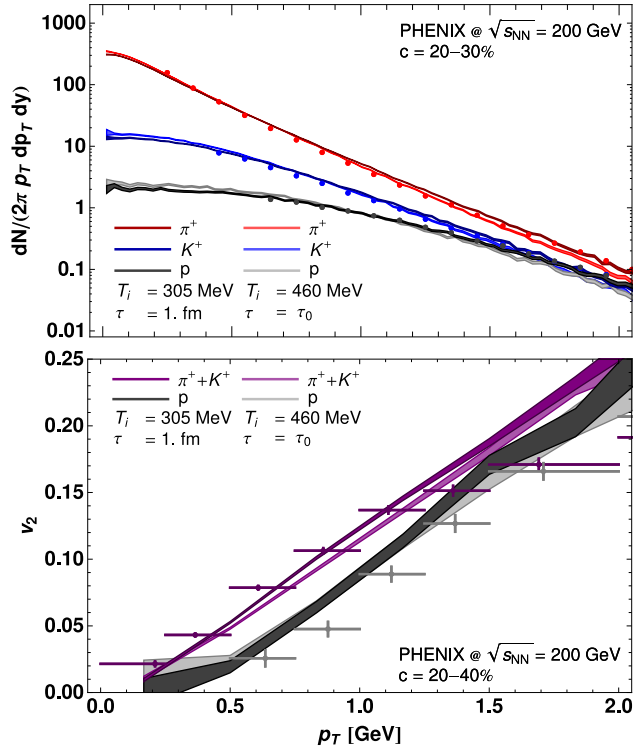}
\end{center}
\vspace{-1mm}
\caption{The transverse-momentum spectra of pions, kaons and protons for the centrality class $c$=20-30\% (upper panel) and the elliptic flow coefficient $v_2$ for  \mbox{$c$=20-40\%} (lower panel). The darker (lighter) lines describe the model results obtained in Ref. \cite{Broniowski:2008qk} for the case with (without) free streaming. Data from \cite{Adler:2003cb,Adler:2003kt}.
\label{fig:spv2}}
\end{figure}

\begin{figure}[tb]
\vspace{2mm}
\begin{center}
\includegraphics[angle=0,width=0.65 \textwidth]{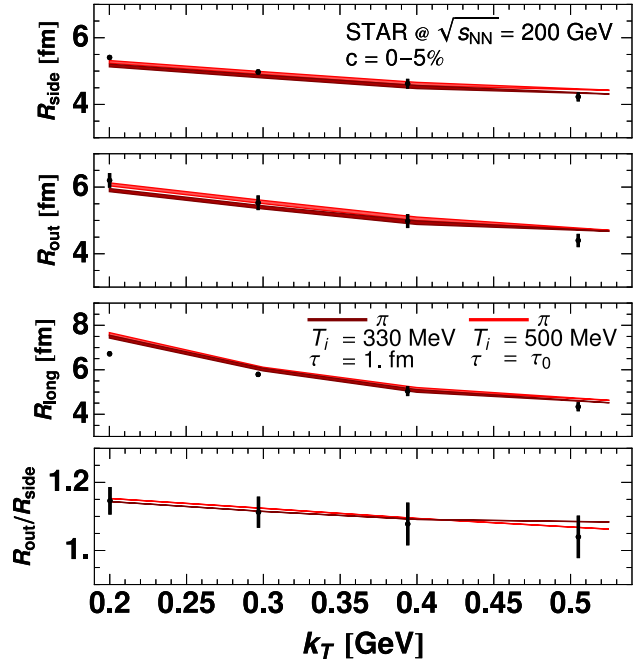}
\end{center}
\vspace{-1mm}
\caption{The pion HBT radii $R_{\rm side}$ , $R_{\rm out}$ , $R_{\rm long}$, and the ratio $R_{\rm out}/R_{\rm side}$ for central collisions. The darker (lighter) lines describe the results obtained in Ref. \cite{Broniowski:2008qk} with (without) FS+SE. The data from \cite{Adams:2004yc}.
\label{fig:hbt}}
\end{figure}

The early hydrodynamic approaches overlooked the fact that for noncentral collisions, where the system develops spatial azimuthal anisotropy, an initial azimuthally asymmetric transverse flow may develop as a consequence of combined free streaming and sudden but delayed equilibration. Interestingly, the results of Ref. \cite{Broniowski:2008qk} obtained with and without the free-streaming stage are practically the same, see Figs. \ref{fig:spv2} and \ref{fig:hbt}. We note that the initial conditions used in \cite{Broniowski:2008qk} are always Gaussian. 

In the case without free streaming, the hydrodynamic stage considered in \cite{Broniowski:2008qk} starts at \mbox{$\tau=\tau_0=0.25$ fm} and the initial central temperature is \mbox{$T_i = 460$ MeV} for \mbox{$c$=20--30\%} and  \mbox{$T_i = 500$ MeV} for \mbox{$c$=0--5\%}. In an alternative scenario, the free-streaming starts at \mbox{$\tau=\tau_0=0.25$ fm} and at \mbox{$\tau=1$ fm} a transition to perfect-fluid hydrodynamics is made. The transition is described with the help of Landau matching conditions. In this case the central temperature at \mbox{$\tau=1$ fm} is \mbox{$T_i = 305$ MeV} for \mbox{$c$=20--30\%} and  \mbox{$T_i = 330$ MeV} for \mbox{$c$=0--5\%}.

\subsection{Early transverse hydrodynamics}
\label{sect:model_2}

The energy-momentum tensor obtained from the free-streaming stage matches very smoothly to the form of the transverse hydrodynamics, where the longitudinal pressure vanishes \cite{Bialas:2007gn,Ryblewski:2008fx}. This behavior  indicates that the free-streaming stage and the perfect-fluid hydrodynamics stage may be separated by a phase that is treated as an anisotropic fluid whose transverse pressure is much larger than the longitudinal pressure. Such pressure anisotropy appears also in the string models \cite{Bialas:1999zg,Florkowski:2003mm} where it is a consequence of the specific production mechanism. 

Having in mind very large pressure anisotropies present at the early stages of heavy-ion collisions, a model was constructed where the initial stage described by the transverse hydrodynamics is matched to the standard perfect-fluid hydrodynamics \cite{Ryblewski:2010tn}. In analogy to the model described in Sect.~\ref{sect:model_1}, the phase described by the transverse hydrodynamics lasts between \mbox{$\tau=0.25$ fm} and  \mbox{$\tau=1$ fm}, and the transition to the perfect-fluid regime is described again by the Landau matching conditions. However, in contrast to the previous model the initial conditions are taken from the Glauber model in the optical approximation.

\begin{figure}[t]
\begin{center}
\includegraphics[angle=0,width=0.65 \textwidth]{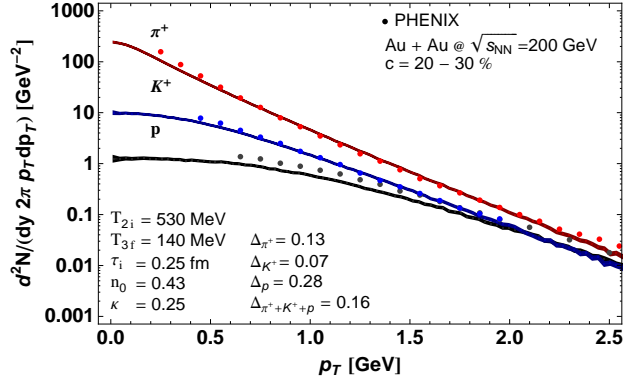}
\end{center}
\vspace{-3mm}
\caption{The comparison of the model \cite{Ryblewski:2010tn} and experimental transverse-momentum spectra of pions, kaons, and protons. The data are taken from Ref. \cite{Adler:2003cb}. The model includes an early stage described by transverse hydrodynamics. The model parameters are defined in \cite{Ryblewski:2010tn}.   }
\label{fig:ppttr}
\end{figure}
\begin{figure}[t]
\begin{center}
\includegraphics[angle=0,width=0.65 \textwidth]{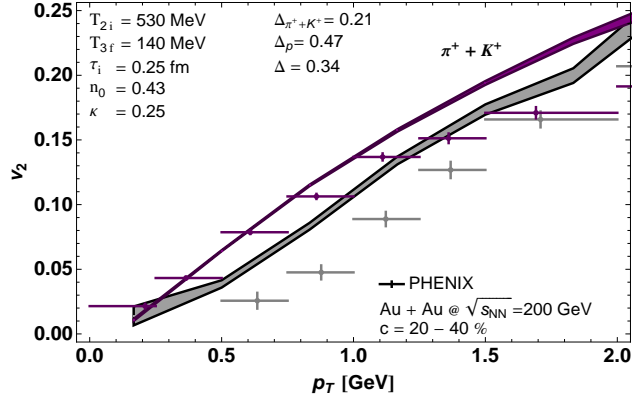}
\end{center}
\vspace{-3mm}
\caption{The comparison of the model \cite{Ryblewski:2010tn} and experimental elliptic flow coefficient $v_2$ for pions+kaons and protons. The data are taken from Ref. \cite{Adler:2003kt}. }
\label{fig:v2tr}
\end{figure}
\begin{figure}[t]
\begin{center}
\includegraphics[angle=0,width=0.65 \textwidth]{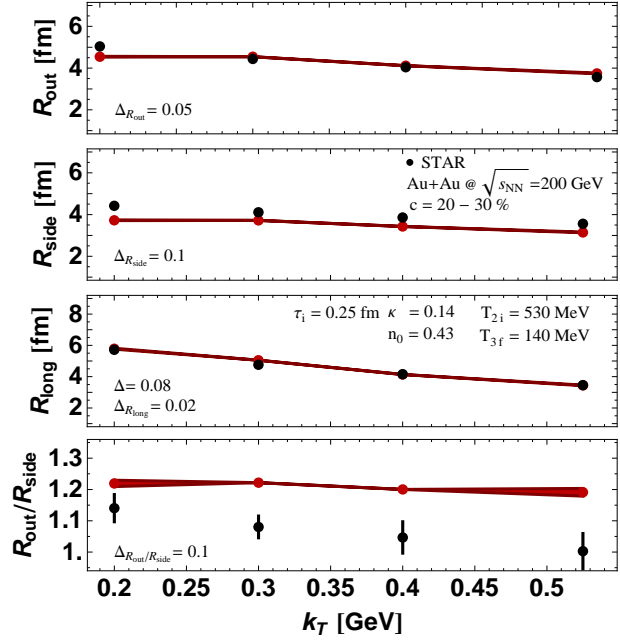}
\end{center}
\vspace{-3mm}
\caption{The comparison of the model \cite{Ryblewski:2010tn} and experimental results for the pionic HBT radii. The data are taken from Ref. \cite{Adams:2004yc}. }
\label{fig:hbttr}
\end{figure}

The model results presented in Figs. \ref{fig:spv2}--\ref{fig:hbt} and \ref{fig:ppttr}--\ref{fig:hbttr}  describe reasonably well the basic soft-hadronic observables measured at RHIC. The model with the transverse-hydrodynamics stage describes the data at the level of 20\% (note the quantity $\Delta$, displayed in Figs. \ref{fig:ppttr}--\ref{fig:hbttr}, which gives the relative difference between the data and the model results). The model with the free-streaming stage describes the data even better (because of the Gaussian initial conditions used for hydrodynamics).

Our two examples of the non-equilibrium early dynamics that precedes the perfect-fluid stage indicate that the early thermalization and the HBT puzzles may be circumvented in the pionic sector; the transverse-momentum spectra, the elliptic flow coefficient $v_2$, and the HBT radii of pions are very well described in the models that do not assume early thermalization. In the two considered cases, the complete thermalization may be postponed to the comfortable time \mbox{$\tau=1$ fm}. This value seems to be acceptable from the point of view of the recent microscopic calculations, see our comments in the end of Sect. \ref{sect:ET}.  

Of the special interest is the fact that the elliptic flow of the pions is reproduced very well together with other observables.  This result indicates that the large pion elliptic flow is not a good signature of very early thermalization.

The free-streaming stage as well as the transverse-hydrodynamics stage change into the phase treated as perfect fluid. In Refs. \cite{Ryblewski:2009hm,Florkowski:2009wb} a physical scenario was considered where the transverse hydrodynamics was followed directly by a sudden isotropization and freeze-out (without the extended perfect-fluid stage). The results of Refs. \cite{Ryblewski:2009hm,Florkowski:2009wb} are similar to those presented in Figs. \ref{fig:ppttr}--\ref{fig:hbttr}. The approach with sudden isotropization and freeze-out gives slightly smaller HBT radii, on the other hand, their $k_T$ dependence is reproduced better, yielding a very good description of the ratio $R_{\rm out}/R_{\rm side}$. The similarities between the results presented in this paper and the results of Refs. \cite{Ryblewski:2009hm,Florkowski:2009wb} indicate that the details of modeling the equilibration transition may be not important. 

\section{HBT or $v_2$ puzzle?}
\label{sect:v2puzzle}

The largest discrepancies between the model predictions presented in Sects. \ref{sect:model_1} and \ref{sect:model_2} and the experimental data can be seen for the proton elliptic flow. The latter is probably too large because hadronic interactions in the final state and/or shear viscosity have been neglected. We know, however, that such interactions spoil the agreement with the HBT data. Clearly, some part of the "HBT puzzle" remains if we study pions, kaons, and protons.  

If we consider hydrodynamic evolution governed by the equations of perfect fluid and calculate physical observables at different freeze-out temperatures $T_f$, we find that with lowering of $T_f$ the spectra become flatter, the splitting of the elliptic flow for different hadrons increases, and the HBT radii become too large. For large $T_f$ (\mbox{$T_f > 140-150$ MeV}, early freeze-out) the spectra are reproduced well, there is no $v_2$ splitting, and the HBT radii are too small. On the other hand, at small $T_f$ (\mbox{$T_f < 140-150$ MeV}, late freeze-out) the spectra are too flat, the correct $v_2$ splitting is reproduced, and the HBT radii are too large. This type of behavior is the main source of the difficulties encountered in attempts to obtain a uniform description of soft-hadronic observables at RHIC.

The optimal results are obtained for moderate temperatures, where the spectra are a bit too flat, a small $v_2$ splitting is obtained, and the HBT radii are well reproduced.  The points i)--ix) discussed in Sect. \ref{sect:HBT} improve significantly this type of optimization, with the effects presented in Sects. \ref{sect:model_1} and \ref{sect:model_2}.

A real progress in improving the results of hydrodynamic calculations has been recently made by P. Bozek who included the effects of bulk viscosity. The inclusion of bulk viscosity lowers the pressure, makes the evolution longer, and leads to the $v_2$ mass splitting. At the same time, the other hydrodynamic predictions describe well the spectra and the HBT radii \cite{Bozek:2009dw}. Another recent calculation by P. Bozek shows that the early thermalization is required to reproduce well the directed flow \cite{Bozek:2010aj}. This may suggest a shift of the ten-years-old paradigm: the directed flow rather than the elliptic flow is a signature of early thermalization. 

\section{Conclusions}

In this paper we have tried to summarize recent attempts to solve the early thermalization and HBT puzzles at RHIC. Many improvements done in modeling of the QCD equations of state, hydrodynamic initial conditions, as well as inclusion of dissipative effects improved the agreement between the hydrodynamic predictions and the data. Such improvements eliminate  most of the discrepancies connected with the notion of the HBT puzzle. The combinations of inviscid hydrodynamics with non-equilibrium models of early stages suggest that the early thermalization puzzle connected with the elliptic flow may be circumvented. However, the new analysis of the directed flow \cite{Bozek:2010aj} shed completely new light on this problem. 

\section{Acknowledgments}

It was my great pleasure to participate in the Jubilee 50th Cracow School of Theoretical Physics in Zakopane. I thank the Organizers for the invitation and providing excellent scientific atmosphere. For the first time I participated in the School in 1984 as a student of physics at the Jagellonian University. During that School Larry McLerran gave the talks about the quark-gluon plasma, a subject which has shaped my scientific activity for many years.

 The results presented in Sects. \ref{sect:model_1} and \ref{sect:model_2} were obtained in collaboration with W. Broniowski, M. Chojnacki, A. Kisiel, and R. Ryblewski. I am grateful to my colleagues for very fruitful collaboration and many illuminating discussions. I also thank P. Bozek for clarifying discussions concerning his newest results. 

This research was supported in part by the Polish Ministry of Science and Higher Education, grant  No. N N202 263438.

%\bibliography{referencje}

\begin{thebibliography}{10}
\expandafter\ifx\csname url\endcsname\relax
  \def\url#1{\texttt{#1}}\fi
\expandafter\ifx\csname urlprefix\endcsname\relax\def\urlprefix{URL }\fi

\bibitem{Heinz:2002un}
U.~W. Heinz, P.~F. Kolb, hep-ph/0204061.

\bibitem{Hirano:2001eu}
T.~Hirano, Phys. Rev., {\bf C65} (2002) 011901.

\bibitem{Heinz:2001xi}
U.~W. Heinz, P.~F. Kolb, Nucl. Phys., {\bf A702} (2002) 269.

\bibitem{Huovinen:2001cy}
P.~Huovinen, P.~F. Kolb, U.~W. Heinz, P.~V. Ruuskanen, S.~A. Voloshin, Phys.
  Lett., {\bf B503} (2001) 58.

\bibitem{Teaney:2001av}
D.~Teaney, J.~Lauret, E.~V. Shuryak, nucl-th/0110037.

\bibitem{Kolb:2002ve}
P.~F. Kolb, R.~Rapp, Phys. Rev., {\bf C67} (2003) 044903.

\bibitem{Gyulassy:2004zy}
M.~Gyulassy, L.~McLerran, Nucl. Phys., {\bf A750} (2005) 30.

\bibitem{Shuryak:2004cy}
E.~V. Shuryak, Nucl. Phys., {\bf A750} (2005) 64.

\bibitem{HanburyBrown:1954wr}
R.~Hanbury~Brown, R.~Q. Twiss, Phil. Mag., {\bf 45} (1954) 663.

\bibitem{HanburyBrown:1956pf}
R.~Hanbury~Brown, R.~Q. Twiss, Nature, {\bf 178} (1956) 1046.

\bibitem{Kopylov:1975rp}
G.~I. Kopylov, M.~I. Podgoretsky, Zh. Eksp. Teor. Fiz., {\bf 69} (1975)
  414.

\bibitem{Baym:1997ce}
G.~Baym, Acta Phys. Polon., {\bf B29} (1998) 1839.

\bibitem{Lednicky:2003mq}
R.~Lednicky, Phys. Atom. Nucl., {\bf 67} (2004) 72.

\bibitem{Makhlin:1987gm}
A.~N. Makhlin, Y.~M. Sinyukov, Z. Phys., {\bf C39} (1988) 69.

\bibitem{Broniowski:2002wp}
W.~Broniowski, A.~Baran, W.~Florkowski, AIP Conf. Proc., {\bf 660} (2003)
  185.

\bibitem{Retiere:2003kf}
F.~Retiere, M.~A. Lisa, Phys. Rev., {\bf C70} (2004) 044907.

\bibitem{Lisa:2005dd}
M.~A. Lisa, S.~Pratt, R.~Soltz, U.~Wiedemann, Ann. Rev. Nucl. Part. Sci., {\bf
  55} (2005) 357.

\bibitem{Lisa:2008gf}
M.~A. Lisa, S.~Pratt, arXiv:0811.1352.

\bibitem{Broniowski:2008vp}
W.~Broniowski, M.~Chojnacki, W.~Florkowski, A.~Kisiel, Phys. Rev. Lett., {\bf
  101} (2008) 022301.

\bibitem{Pratt:2008qv}
S.~Pratt, Phys. Rev. Lett., {\bf 102} (2009) 232301.

\bibitem{Chojnacki:2007jc}
M.~Chojnacki, W.~Florkowski, Acta Phys. Polon., {\bf B38} (2007) 3249.

\bibitem{Huovinen:2009yb}
P.~Huovinen, P.~Petreczky, Nucl. Phys., {\bf A837} (2010) 26.

\bibitem{Bozek:2009ty}
P.~Bozek, I.~Wyskiel, Phys. Rev., {\bf C79} (2009) 044916.

\bibitem{Chojnacki:2004ec}
M.~Chojnacki, W.~Florkowski, T.~Csorgo, Phys. Rev., {\bf C71} (2005) 044902.

\bibitem{Sinyukov:2009xe}
Y.~M. Sinyukov, A.~N. Nazarenko, I.~A. Karpenko, Acta Phys. Polon., {\bf B40}
  (2009) 1109.

\bibitem{Socolowski:2004hw}
O.~Socolowski, Jr., F.~Grassi, Y.~Hama, T.~Kodama, Phys. Rev. Lett., {\bf 93}
  (2004) 182301.

\bibitem{Kisiel:2006is}
A.~Kisiel, W.~Florkowski, W.~Broniowski, Phys. Rev., {\bf C73} (2006) 064902.

\bibitem{Broniowski:2001we}
W.~Broniowski, W.~Florkowski, Phys. Rev. Lett., {\bf 87} (2001) 272302.

\bibitem{Teaney:2003kp}
D.~Teaney, Phys. Rev., {\bf C68} (2003) 034913.

\bibitem{Romatschke:2007mq}
P.~Romatschke, U.~Romatschke, Phys. Rev. Lett., {\bf 99} (2007) 172301.

\bibitem{Pratt:2008sz}
S.~Pratt, J.~Vredevoogd, Phys. Rev., {\bf C78} (2008) 054906.

\bibitem{Andrade:2006yh}
R.~Andrade, F.~Grassi, Y.~Hama, T.~Kodama, O.~Socolowski, Jr., Phys. Rev.
  Lett., {\bf 97} (2006) 202302.

\bibitem{Broniowski:2007nz}
W.~Broniowski, M.~Rybczynski, P.~Bozek, Comput. Phys. Commun., {\bf 180} (2009)
  69.

\bibitem{Werner:2010aa}
K.~Werner, I.~Karpenko, T.~Pierog, M.~Bleicher, K.~Mikhailov, arXiv:1004.0805.

\bibitem{Schenke:2010rr}
B.~Schenke, S.~Jeon, C.~Gale, arXiv:1009.3244.

\bibitem{Kisiel:2005hn}
A.~Kisiel, T.~Taluc, W.~Broniowski, W.~Florkowski, Comput. Phys. Commun., {\bf
  174} (2006) 669.

\bibitem{Bowler:1991vx}
M.~G. Bowler, Phys. Lett., {\bf B270} (1991) 69.

\bibitem{Sinyukov:1998fc}
Y.~Sinyukov, R.~Lednicky, S.~V. Akkelin, J.~Pluta, B.~Erazmus, Phys. Lett.,
  {\bf B432} (1998) 248.

\bibitem{Maj:2009ue}
R.~Maj, S.~Mrowczynski, Phys. Rev., {\bf C80} (2009) 034907.

\bibitem{Broniowski:2008qk}
W.~Broniowski, W.~Florkowski, M.~Chojnacki, A.~Kisiel, Phys. Rev., {\bf C80}
  (2009) 034902.

\bibitem{Torrieri:2004zz}
G.~Torrieri, {\it et~al.}, Comput. Phys. Commun., {\bf 167} (2005) 229.

\bibitem{Geiger:1994he}
K.~Geiger, Phys. Rept., {\bf 258} (1995) 237.

\bibitem{Xu:2007qn}
Z.~Xu, C.~Greiner, Eur. Phys. J., {\bf C49} (2007) 187.

\bibitem{Xu:2004gw}
X.-M. Xu, Y.~Sun, A.-Q. Chen, L.~Zheng, Nucl. Phys., {\bf A744} (2004)
  347.

\bibitem{Baier:2000sb}
R.~Baier, A.~H. Mueller, D.~Schiff, D.~T. Son, Phys. Lett., {\bf B502} (2001)
  51.

\bibitem{Baier:2002bt}
R.~Baier, A.~H. Mueller, D.~Schiff, D.~T. Son, Phys. Lett., {\bf B539} (2002)
  46.

\bibitem{El:2007vg}
A.~El, Z.~Xu, C.~Greiner, Nucl. Phys., {\bf A806} (2008) 287.

\bibitem{Mrowczynski:1994xv}
S.~Mrowczynski, Phys. Rev., {\bf C49} (1994) 2191.

\bibitem{Arnold:2004ti}
P.~Arnold, J.~Lenaghan, G.~D. Moore, L.~G. Yaffe, Phys. Rev. Lett., {\bf 94}
  (2005) 072302.

\bibitem{Bialas:1999zg}
A.~Bialas, Phys. Lett., {\bf B466} (1999) 301.

\bibitem{Florkowski:2003mm}
W.~Florkowski, Acta Phys. Polon., {\bf B35} (2004) 799.

\bibitem{Kharzeev:2005iz}
D.~Kharzeev, K.~Tuchin, Nucl. Phys., {\bf A753} (2005) 316.

\bibitem{Castorina:2007eb}
P.~Castorina, D.~Kharzeev, H.~Satz, Eur. Phys. J., {\bf C52} (2007) 187.

\bibitem{Biro:1993qc}
T.~S. Biro, C.~Gong, B.~Muller, A.~Trayanov, Int. J. Mod. Phys., {\bf C5}
  (1994) 113.

\bibitem{Mrowczynski:2005ki}
S.~Mrowczynski, Acta Phys. Polon., {\bf B37} (2006) 427.

\bibitem{Kolb:2000sd}
P.~F. Kolb, J.~Sollfrank, U.~W. Heinz, Phys. Rev., {\bf C62} (2000) 054909.

\bibitem{Jas:2007rw}
W.~Jas, S.~Mrowczynski, Phys. Rev., {\bf C76} (2007) 044905.

\bibitem{Sinyukov:2006dw}
Y.~M. Sinyukov, Acta Phys. Polon., {\bf B37} (2006) 3343.

\bibitem{Gyulassy:2007zz}
M.~Gyulassy, Y.~M. Sinyukov, I.~Karpenko, A.~V. Nazarenko, Braz. J. Phys., {\bf
  37} (2007) 1031.

\bibitem{Adler:2003cb}
S.~S. Adler, {\it et~al.}, PHENIX, Phys. Rev., {\bf C69} (2004) 034909.

\bibitem{Adler:2003kt}
S.~S. Adler, {\it et~al.}, PHENIX, Phys. Rev. Lett., {\bf 91} (2003) 182301.

\bibitem{Adams:2004yc}
J.~Adams, {\it et~al.}, STAR, Phys. Rev., {\bf C71} (2005) 044906.

\bibitem{Bialas:2007gn}
A.~Bialas, M.~Chojnacki, W.~Florkowski, Phys. Lett., {\bf B661} (2008)
  325.

\bibitem{Ryblewski:2008fx}
R.~Ryblewski, W.~Florkowski, Phys. Rev., {\bf C77} (2008) 064906.

\bibitem{Ryblewski:2010tn}
R.~Ryblewski, W.~Florkowski, Phys. Rev., {\bf C82} (2010) 024903.

\bibitem{Ryblewski:2009hm}
R.~Ryblewski, W.~Florkowski, Acta Phys. Polon. Supp., {\bf B3} (2010) 557.

\bibitem{Florkowski:2009wb}
W.~Florkowski, R.~Ryblewski, J. Phys., {\bf G37} (2010) 094023.

\bibitem{Bozek:2009dw}
P.~Bozek, Phys. Rev., {\bf C81} (2010) 034909.

\bibitem{Bozek:2010aj}
P.~Bozek, I.~Wyskiel-Piekarska, arXiv:1009.0701.

\end{thebibliography}

\end{document}